\documentclass[aps,pra,nofootinbib,groupedaddress,twocolumn,final]{revtex4}
\begin{document}

\def\sb{{\sf b}}
\def\cG{{\cal G}}
\def\cH{{\cal H}}
\def\cK{{\cal K}}
\def\cQ{{\cal Q}}
\def\tr{{\rm tr}}
\def\AA{{\rm A}}
\def\BB{{\rm B}}
\def\CC{{\rm C}}
\def\UU{{\rm U}}
\def\VV{{\rm V}}

\newcommand\ket[1]{| #1 \rangle}
\newcommand\bra[1]{\langle #1 |}
\newcommand\braket[2]{\langle #1|#2\rangle}
\newcommand\ketbra[2]{|#1\rangle\langle #2|}
\newcommand\norm[1]{\| #1 \|}
\newcommand\trnorm[1]{\| #1 \|_1}

\def\drangle{{\leavevmode{\rm \rangle\mkern -4mu \rangle}}}
\def\dlangle{{\leavevmode{\rm \langle\mkern -4mu \langle}}}

\newcommand{\dket}[1]{| #1 \drangle}
\newcommand{\dbra}[1]{\dlangle #1 |}
\newcommand{\dbraket}[2]{\dlangle #1 | #2 \drangle}
\newcommand{\dketbra}[2]{| #1 \drangle \dlangle #2 |}

\title{Another simple unconditionally secure quantum
bit commitment protocol --- beating entanglement with entanglement\footnote{NOTE: This paper is the second of three papers
that together provide a detailed description of various
  gaps in the QBC ``impossibility proof,'' as well as security
  proofs for four different protocols, QBC1, QBC2, QBC4, and QBC5. In this v3 on
  QBC4, the protocol is modified to remove an ambiguity in v2 in a
  more transparent formulation. The appendix has been moved to the
  first paper.}}
\author{Horace P. Yuen}\email{yuen@ece.northwestern.edu}
\affiliation{Department of Electrical and Computer Engineering,
  Department of Physics and Astronomy, Northwestern University,
  Evanston, IL 60208}
\begin{abstract}
It is shown how the evidence state space in quantum bit commitment may
be made to depend on the bit value 0 or 1 with split entangled
pairs. As a consequence, one can obtain a protocol that is perfectly
concealing, but is also $\epsilon$-binding because the bit-value
dependent evidence space prevents the committing party from cheating
by means of a local transformation that is independent of the part of
evidence state space that has never been in his possession.
\end{abstract}
\maketitle

In the previous paper \cite{yuen1}, we showed how quantum teleportation may be
used to perform unconditionally secure quantum bit commitment (US
QBC) by preventing the committing party (Adam) from entangling the
different commitment possibilities. In this paper, we will show that
another simple US QBC protocol may be obtained by individuating the
evidence state space via split entangled pairs, so that pefect
concealing can be achieved, while Adam is faced with a bit-value
dependent evidence space. This in turn ensures that Adam cannot cheat
by means of a local transformation. The underlying idea can be
explained as follows. In order to cheat successfully, Adam needs to
know the total state including the part Babe always keeps that does not
affect security. However, Babe can classically randomize over that part
and she cannot cheat by entangling over it. This interplay of classical and
quantum randomness can be utilized to yield a simple US QBC
protocol as follows.

Let $\ket{k_j}_j$, $j \in \{ \mu,\nu\}$, $k_j \in \{1,2\}$, be two openly known orthonormal qubit
states, $\braket{1}{2}=0$, for each of the two possible $j$. When
there is no ambiguity, we would write $\ket{k_j}_j$ simply as
$\ket{k}_j$ to simplify notation. Let Babe prepare two states
\begin{equation}
\ket{\Psi_j} = \frac{1}{\sqrt{2}} \sum_k \ket{k}_j \ket{f_k}_j,
\label{eq:psij}
\end{equation}
where $\ket{k}_j \in \cH^B_{j\alpha}$, $k \in \{1,2\}$, and
$\{\ket{f_k}_j | k=1,2\}$ form an orthonormal basis in
$\cH^B_{j\beta}$ for each $j \in \{\mu,\nu\}$, with $\ket{\Psi_j} \in
\cH^B_{j\alpha} \otimes \cH^B_{j\beta}$ on two qubits for each
$j$. We have skipped one subscript $j$ in $\ket{f_{k_j}}_j$ as in
$\ket{k}_j$ to simplify notation. Let $\cH^B_\alpha \equiv
\cH^B_{\mu\alpha} \otimes \cH^B_{\nu\alpha}$, $\cH^B_\beta \equiv
\cH^B_{\mu\beta} \otimes \cH^B_{\nu\beta}$, $\cH^B \equiv \cH^B_\alpha
\otimes \cH^B_\beta$.

Babe keeps $\cH^B_\beta$ and sends the ordered pair of qubits
$\cH^B_\alpha$ to Adam. Adam applies the followin transformation on
$\cH^B_{j\alpha}$ separately for each $j$: $\ket{\Psi_j}$ becomes
$\ket{\Phi_j} \in \cH^A_j \otimes \cH^B_{j\alpha} \otimes
\cH^B_{j\beta}$:
\begin{equation}
\ket{\Phi_j} = \frac{1}{\sqrt{8}} \sum_{k,i} \ket{e_i}_j V_i \ket{k}_j
\ket{f_k}_j,
\label{eq:phij}
\end{equation}
where $i \in \{1,2,3,4\}$, $\{\ket{e_i}_j\}$ complete orthonormal in
$\cH^A_j$, and $V_i$ are four unitary qubit operators given by $I$,
$\sigma_x$, $-i\sigma_y$, $\sigma_z$ in terms of the Pauli spin
operators when $\ket{1}$ and $\ket{2}$ lie on the qubit
$z$-axis. Eq.~(\ref{eq:phij}) can be obtained by the unitary
transformation $\sum_i \ketbra{e_i}{e_i} \otimes V_i$ on $\cH^A
\otimes \cH^B_{j\alpha}$ with initial state $\ket{\psi_A} \in \cH^A$
that has $\braket{e_i}{\psi_A} = \frac{1}{2}$. To commit $\sb=0$, Adam
sends back $\cH^B_{\mu\alpha} \otimes \cH^B_{\nu\alpha}$ in the
original order, and he switches them to $\cH^B_{\nu\alpha}\otimes
\cH^B_{\mu\alpha}$ to commit $\sb=1$. He opens by announcing $\sb$,
the order of the two $\cH^B_{j\alpha}$ he committed, and submitting
the ordered qubit pair $\cH^A \equiv \cH^A_\mu \otimes
\cH^A_\nu$. Babe verifies by measuring the corresponding projections
to $\ket{\Phi_\mu}\ket{\Phi_\nu}$ of (\ref{eq:phij}).

It is easy to verify by tracing over $\cH^A$ that for either $\sb$,
$\rho^B_0 = \rho^B_1 = I^B/16$ on $\cH^B$, for any orthonormal
$\{\ket{f_k}_j\}$. If Babe entangles over the possible choices of such
$\{ \ket{f_k}_j\}$, a simple calculation shows that perfect concealing
$\rho^{BC}_0 = \rho^{BC}_1$ on $\cH^B \otimes \cH^C$ is maintained,
where $\cH^C$ is the space Babe used to carry out such
entanglement. This happens because the $V_i$ operations by Adam
totally disentangle the state on $\cH^B_\alpha \otimes \cH^B_\beta
\otimes \cH^C$ into a product state $I^B_\alpha / 4 \otimes
\rho^{BC}_\beta$ for either $\sb$, and there is no identity that
individuates a qubit that is not entangled to another with both qubits
in
one's possession.

Intuitively, we intend to guarantee binding by the fact that
$\cH^B_{j\beta} = \cH^B_{\mu\beta} \otimes \cH^B_{\nu\beta}$ in Babe's
possession cannot be switched to $\cH^B_{\nu\beta}\otimes
\cH^B_{\mu\beta}$ by operating on $\cH^A \otimes \cH^B_\alpha$
alone. However, this is possible if the two orthonormal sets $\{
\ket{f_k}_j\}$ are known. Indeed, this is the content of the
impossibility proof \cite{note}. Thus, to guarantee security, Babe
needs to employ different choices of $\{\ket{f^n_k}_j\}$ with
different bases indexed by $n$. She may employ a fixed probability
distribution $\{ p_{nj}\}$ for each $j$, and she may entangle these
via orthonormal $\{ \ket{g^n}_j\}$, ad infinitum. This possible chain
of purifications has to stop somewhere, and we simply stop it at
$\cH^B$ without $\cH^C$. As we have seen, this does not affect perfect
concealing so that Babe is free to choose any orthonormal
$\{\ket{f_k}_j\}$. It is clearly unreasonable for Adam to demand such
knowledge, as codified in the Secrecy Principle of Ref.~\cite{yuen2}
in our discussion of what we call Type 3 protocols. This possibility
is neglected in the impossibility proof. In Ref.~\cite{yuen3}, a proof
was given that the knowledge of entanglement basis by Babe is not
needed for Adam's cheating for a class of protocols that do not
involve the switching of evidence state for commitment or the
submission of part of an entangled state by Adam upon opening. It is
these two features in combination that guarantee the security of the
present protocol. Further elaboration is given in connection with Type
3 protocols in Ref.~\cite{yuen4}.

To see exactly how binding is obtained in the present situation, note
that the perfect cheating transformation $U^A$ is determiend by
Eq.~(18) of Ref.~\cite{yuen3}, which is unique up to a phase factor in
this nondegenerate situation. It depends on the unitary matrix
$V_{kk'} ={}_\mu \braket{f_k}{f_{k'}}_\nu$ in the present case with
state-space switching, in contrast to merely $\braket{f_k}{f_{k'}} =
\delta_{kk'}$, i.e., no dependence on the actual $\{\ket{f_k}\}$ in
the case without switching. Thus, Adam
cannot cheat perfectly. Note that we are indeed beating
entanglement with entanglement: Babe's entanglement in the form
(\ref{eq:psij}) is essential. She cannot maintain the protocol
security against Adam by just sending $\cH^B_{\mu\alpha} \otimes
\cH^B_{\nu\alpha}$ to Adam without first entangling to
$\cH^B_\beta$. On the other hand, Adam's entanglement is not
essential. As usual in QBC protocols, the whole procedure works the
same if Adam chooses the $V_i$ on $\cH^B_{\mu\alpha}$ and
$\cH^B_{\nu\alpha}$ classically and opens by telling Babe his choice.

We have assumed as usual that Adam opens $\sb=0$ perfectly. Let $p_A <
1$ be Adam's optimum probability of cheating for a given choice of $\{ \ket{f^n_k}_j\}$ and $\{
p_{nj}\}$, taking into account also all his other obvious imperfect
cheating possibilities, such as simply announcing a different
$\sb$. We have thus shown that the formulation and the reasoning of
the impossibility proof break down already in this simple pair
$\ket{\Phi_\mu}\ket{\Phi_\nu}$ situation.

When $\sb=0$ pefect opening condition is relaxed, it is clear that
Adam still cannot cheat perfectly, but it is possible that the overall
successful opening probability (honest plus cheating) may be improved. By continuity it can be seen that Adam's optimum
cheating probability $\bar{P}^A_c$ is arbitrarily close to $p_A=\frac{1}{2}$ if
the $\sb=0$ opening probability is arbitrarily close to $1$, the case
of interest.

Protocol QBC4 is obtained when the above protocol, to be called QBC4p, is extended to a sequence of $\{
\ket{\Psi_{\ell \mu}}\ket{\Psi_{\ell \nu}}\}$, $\ell \in \{1,\ldots,N\}$,
each of the form (\ref{eq:psij}), with $\ket{f_{\ell k}}_j \in
\cH^B_{\ell j \beta}$, $\ket{k_\ell}_j \in \cH^B_{\ell j \beta}$,
etc. Babe should send Adam $\{ \cH^B_{\ell \mu \alpha} \otimes \cH^B_{\ell
\nu \alpha}\}$ and Adam should commit to Babe these spaces for all
$\mu$ after he entangles them with $\cH^A_{\ell \mu} \otimes
\cH^A_{\ell \nu}$ using the $V_i$ operations, permuting each pair for
$\sb=1$. He opens by announcing $\sb$ and the state of the qubits in
each $\cH^B_{\ell \alpha}$ and
submitting $\{\cH^A_\ell\}$, with Babe verifyng
$\ket{\Phi_{\ell\mu}}\ket{\Phi_{\ell\nu}} \in \cH^A_\ell \otimes
\cH^B_\ell$ after possible rearrangement for each $\ell$. Since there
is no new entanglement possibility for Adam, the protocol is perfectly
concealing with $\bar{P}^A_c = p^N_A$ going to zero exponentially in
$N$. Thus, QBC4 is perfectly concealing and $\epsilon$-binding for any
$\epsilon > 0$ by letting $N$ be large.

So far we have assumed Babe is honest in sending Adam $\{ \cH^B_{\ell\mu\alpha} \otimes \cH^B_{\ell\nu\alpha}\}$ with states $\{
\ket{\Psi_{\ell \mu}} \ket{\Psi_{\ell \nu}} \}$. However, she could cheat
by sending in different states, e.g., unentangled states which are
orthogonal for $\mu$ and $\nu$. This kind of cheating is not accounted
for in the impossibility proof formulation, which assumes the parties are honest during
commitment, but can be handled in an ensemble formulation or a game-theoretic
formulation as quantitatively described in Appendices A and B of
Ref.~\cite{yuen1}. In this case, Adam checks
$\ket{\Psi_\mu}\ket{\Psi_\nu}$ each time by asking Babe to send him
$\cH^B_\beta$ and check that the state in $\cH^B_\alpha \otimes
\cH^B_\beta$ is of the form (\ref{eq:psij}) for some $\{\ket{f_k}_j\}$. We summarize our perfectly concealing and
$\epsilon$-binding protocol:

\begin{center}
\vskip 0.12in
\framebox{
\begin{minipage}{0.9\columnwidth}
\vskip 0.1in
\underline{PROTOCOL {\bf QBC4}}

{\small \begin{enumerate}
\item[(i)] Babe sends Adam $N$ ordered pairs $\{ \cH^B_{\ell \mu\alpha}
\otimes \cH^B_{\ell \nu\alpha} \}$ of qubit pairs, $\ell \in \{1,\ldots,N\}$,
which are entangled to $\{ \cH^B_{\ell \mu\beta} \otimes \cH^B_{\ell \nu\beta}\}$
in her possession in states $\ket{\Psi_{\ell \mu}}\ket{\Psi_{\ell \nu}}$
of the form (\ref{eq:psij}), with independent random choices of
$\{\ket{f^n_k}_j\}$ with probability $\{p_{nj}\}$.
\item[(ii)] To commit $\sb$, Adam applies, for each $\ell$, $\sum_i
\ketbra{e_i}{e_i} \otimes V_i$ on $\cH^A_\ell \otimes \cH^B_{\ell
  \alpha}$, resulting in a state
$\ket{\Phi_{\ell\mu}}\ket{\Phi_{\ell\nu}}$ given via
the form (\ref{eq:phij}), and sends $\{ \cH^B_{\ell \alpha}\}$ to
Babe as evidence for $\sb=0$, while switching the order ot each
$\cH^B_{\ell\mu\alpha}\otimes \cH^B_{\ell\nu\alpha}$ for $\sb=1$.
\item[(iii)] Adam opens by announcing $\sb$, the order of the qubits
  in each $\cH^B_{\ell\alpha}$, and submitting $\{
\cH^A_\ell\}$. Babe verifies by projective measurements of $\{
\ket{\Phi_{\ell\mu}}\}$, $\{\ket{\Phi_{\ell\nu}}\}$, for all $\ell$.
\end{enumerate}
\vskip 0.1in
}
\end{minipage}
}
\end{center}
\vskip 0.15in

This protocol belongs to what we call Type 4 protocols, in which split-entangled pairs are used to individuate state spaces
$\cH^B_{\mu\alpha}$ and $\cH^B_{\nu\alpha}$ for verification, while
they are indistinguishable to Babe before opening. In this way, both perfect concealing and
$\epsilon$-binding can be obtained in a situation not covered by the
impossibility proof. This protocol also utilizes the essential feature
of a Type 3 protocol, as discussed in Ref.~\cite{yuen4}. There we
would also describe another protocol QBC2, in which the switching of
evidence state spaces is employed without split-entangled pairs, but
only with a resulting $\epsilon$-concealing protocol.

\begin{acknowledgments}
I would like to thank G.M.~D'Ariano, W.Y.~Hwang, H.K.~Lo, and R.~Nair for
useful discussions. This work was supported by the Defence Advanced Research Project
Agency and The Army Research Office.
\end{acknowledgments}


\begin{thebibliography}{99}
\bibitem{yuen3}H.P. Yuen, quant-ph/0109055. Note that definition of
  $\Lambda_{ij}$ given by Eq.~(19) is actually that of $\Lambda_{ji}$,
  as can be seen from Appendix A of that paper.
\end{thebibliography}
\end{document}